Costantino Sigismondi
*Università di Roma "La Sapienza"*
sigismondi@icra.it

# Occultazione asteroidale di 474 Prudentia su HIP 1927 con pre-pointing

**Abstract:** An asteroidal occultation of a 8.5 V magnitude star can be difficult to be observed without an automatic pointing system. It was the case of 474 Prudentia occulting HIP 1927 on 5 September 2010, observed from the center of Rome.

The solution of pre-pointing the telescope to Jupiter and leaving it fixed for 20 m 30 s, allowed to have the correct field of view and to observe successfully the phenomenon. This was possible because Jupiter was at the same declination of the target star, and 20m30s of right ascension to the west of it.

With opportune modifications this method can be applied in similar cases, when the observing equipment is minimal. In that case, for example, many observers could have been witness of such an event, helping to recover the shadow path across the city.

### Introduzione: l'osservazione delle occultazioni asteroidali

La prima occultazione asteroidale osservata fu quella del grande asteroide 3 Juno, nel 1958 in Svezia, ma solo a partire dagli anni '70 fu possibile fare delle previsioni sufficientemente accurate su questi eventi. Oggi sono diversi gli studiosi che si occupano dell'astrometria degli asteroidi, sia nelle previsioni, che nella riduzione dei dati, e molti di piú nell'osservazione.

Ad esempio l´8 Luglio 2010 delta Ophiuchi, una stella di seconda grandezza, è stata occultata da parte dell'asteroide 472 Roma, data la buona stagione ed il percorso dell'ombra su mezza Europa, e circa 150 osservatori sono stati coinvolti, ma molti di essi andati sulla prevista centerline, non l'hanno osservata, a causa degli errori nelle effemeridi.

L'esistenza di questi errori dimostra che l´osservazione delle occultazioni asteroidali puó contribuire significativamente al miglioramento delle effemeridi, oltre che allo studio della morfologia di questi corpi minori del sistema solare.

Nel video di questa occultazione [1] si può notare anche l'effetto del diametro angolare della stella (una supergigante di circa 9 centesimi di secondo d'arco di diametro), per cui la sua occultazione è stata graduale, come è stato osservato per Regolo occultata da 166 Rhodope, di cui l'autore ha pubblicato video e reports scientifici [2,3,4,5]. Nella occultazione della stella HIP 1927 da parte di 474 Prudentia [6], qui descritta, la sparizione é stata istantanea.

La luminosità della stella occultata ed il passaggio dell'ombra previsto su Roma mi ha stimolato a tentare di osservare il fenomeno dal balcone della cucina, nonostante le luci della città ed il telescopio, per dirla con un eufemismo, in "modalità Dobson" ovvero senza montatura equatoriale motorizzata né puntamento automatico (fig. 2). Anche la ricerca del giusto campo stellare con il cercatore e le mappe del cielo risulta particolarmente complicata in questo caso.

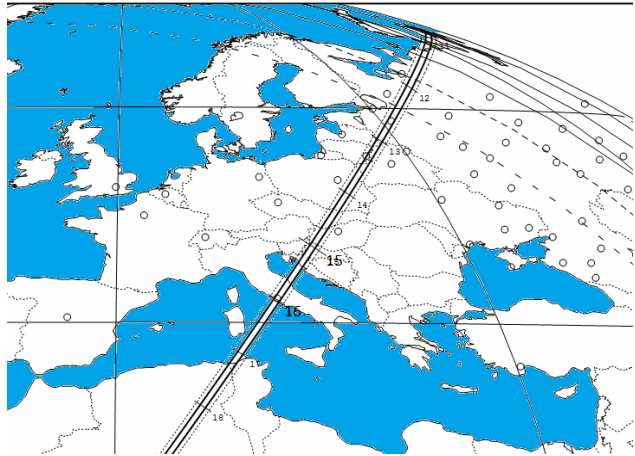

**Fig. 1** Previsione calcolata da Steve Preston del percorso dell'ombra dell'asteroide 474 Prudentia la notte del 5 settembre 2010 sull´Europa.

Tutti i dati riportati nella tabella sopra la figura 1 completano l'informazione [7]: la magnitudine V della stella (8.5), ed anche la R (7.7) e la B (9.9); il diametro dell'asteroide 38 Km e la sua parallasse 8.5" che corrisponde ad una Unitá Astronomica di distanza; il moto orario dell´asteroide pari a 27 arcsec in declinazione e 1.2 s in ascensione retta, corrispondenti a 18 arcsec, complessivamente l'asteroide viaggiava a 32.4 arcsec per ora, cioè 9 milliarcsec al secondo; le dimensioni angolari dell'asteroide 43x28milliarcsec, l'angolo di posizione del suo semiasse maggiore e la sua magnitudine 12.9, assolutamente invisibile all'occhio nudo sia pure con un obbiettivo di 20 cm a causa del cielo di Roma.

L'ombra parte da Nord Est e lascia Roma al termine del 16° minuto (in evidenza) dopo l´1 UT. L'occultazione è stata osservata effettivamente alle 1:15:59 UT.

A Terra la velocitá dell'ombra era di quasi 7 Km/s.[7]

A cosa puó contribuire una osservazione amatoriale di una occultazione asteroidale? L'osservazione di una occultazione ha valore anche se il risultato è di tipo Booleano: "vista o non vista", "si o no", senza fornire il valore esatto della durata dell'occultazione né l'istante di UTC a cui questo fenomeno ha avuto inizio, ma solo la posizione esatta dell'osservatore.

È evidente, dunque, che la cosa piú importante per un'osservazione di questo genere sia l'individuazione della stella target, senza ambiguitá di sorta.

### Il puntamento del telescopio

Il primo problema da affrontare è ritrovare la stella giusta tra tante. La presenza di Giove 5 gradi ad Ovest della stella è stato il primo indizio su come avventurarsi in un campo di vista mai frequentato fino ad allora. L'idea di usare la rotazione della sfera celeste per puntare il telescopio è venuta confrontando le coordinate della stella target e quelle di Giove, per queste ultime ho usato Ephemvga [8], un programma DOS sufficientemente preciso.

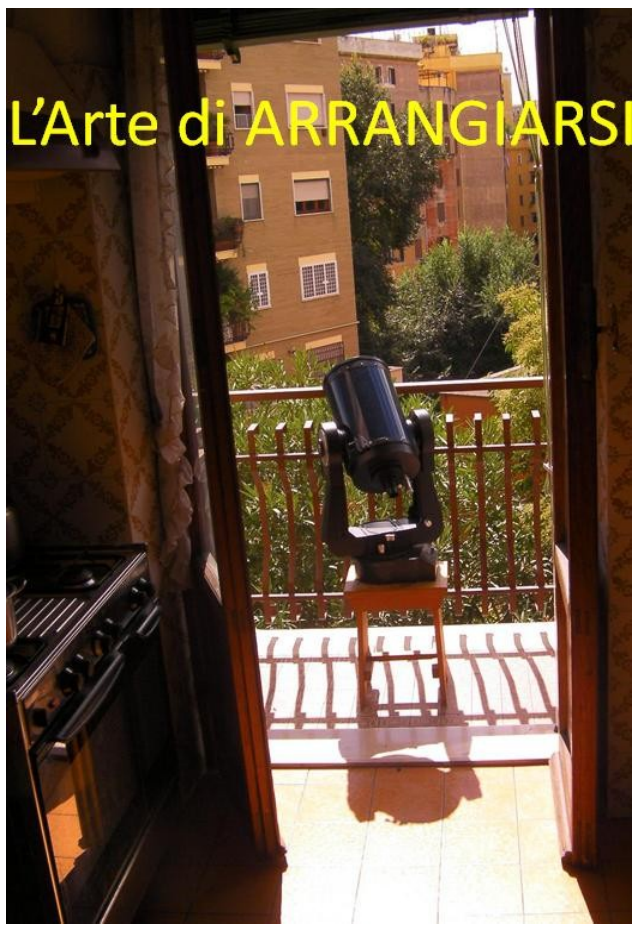

**Fig. 2** Telescopio in "modalità Dobson" e il balcone dell'osservazione con vista Sud. Una curiositá: passa anche su questo balcone il primo meridiano d'Italia 12°27'08" E. Il titolo della figura è un omaggio a Napoli, dove ho presentato questo lavoro nel 43° Congresso dell´UAI dal 23 al 26/9/2010.

La stella, alla stessa declinazione di Giove, è stata individuata con la massima precisione, attendendo 20 m 30 s col telescopio fermo (la differenza tra le ascensioni rette di Giove e della stella) e con il movimento manuale è stata mantenuta nel campo fino al momento dell´occultazione.

**Risultati**

L´osservazione è stata positiva: la stella è sparita per poco meno di un secondo. In mancanza di un video puó essere utile una registrazione audio sincronizzata con un segnale orario standard e si raccomanda di pronunciare un suono secco (es: "è" "tà" ) alla sparizione e riapparizione.

La durata osservata di 1 s di occultazione rispetto alla massima calcolata di 5.5 s indica che le coordinate di casa mia (41°52.8' N 12°27.1' E) sono presso un bordo dell'ombra: rientrano nella fascia Nord dell'ombra calcolata da Preston entro il limite 1, facendo scartare il limite 3 perché troppo lontano per dar luogo ad un'occultazione di 1 s dal mio balcone [7]: questo è già un ottimo risultato.

La mia sola osservazione dell'occultazione di 474 Prudentia su HIP 1927 non consente di sapere, indipendentemente dalle effemeridi, se la mia casa si trovava a destra o a sinistra della centerline, dove l'occultazione avrebbe avuto durata massima. Un altro osservatore a pochi chilometri di distanza misurati perpendicolarmente al percorso dell'ombra avrebbe potuto confermare osservativamente l'ipotesi, indipendentemente dalle effemeridi, correggendole, quindi, o confermandole.

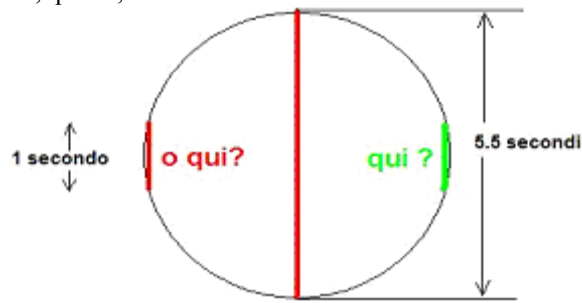

**Fig. 3 Durata dell'occultazione osservata in confronto con quella calcolata per il centro dell'ombra.**

**Conclusioni**

Se l´idea di usare Giove come pre-pointer mi fosse venuta prima, non sarebbe stato difficile coinvolgere altri osservatori e campionare meglio l´ombra di 474 Prudentia sopra Roma.

Il metodo del puntamento di un corpo celeste più luminoso e più facile da individuare, che abbia stessa declinazione e sia poco piú a Ovest, puó facilitare l´osservazione di altre occultazioni asteroidali, anche se l'equipaggiamento non è quello delle migliori occasioni, o nel caso di stazioni remote senza osservatore, metodo già adottato dalla IOTA con i sistemi "mighty mini".[9]

Per una lista delle prossime occultazioni, con le mappe di previsione, consiglio di consultare i siti UAI e IOTA.[10]